\documentclass[openacc]{rstransa}

\begin{document}

\title{Deuterated forms of H${_3^+}$ and their importance in astrochemistry}

\author{
P. Caselli$^{1}$, O. Sipil\"a$^{1}$ and J. Harju$^{2,1}$}

\address{$^{1}$Max-Planck-Institut f\"ur Extraterrestrische Physik, Gie{\ss}enbachstrasse 1, 85741 Garching bei M\"unchen, Germany\\
$^{2}$Department of Physics, P.O. BOX 64, FI-00014 University of Helsinki, Finland\\}

\subject{Astrophysics, Interstellar Medium, Nuclear Atomic and Molecular Physics}


\corres{Insert corresponding author name\\
\email{caselli@mpe.mpg.de}}

\begin{abstract}
At the low temperatures ($\sim$10 K) and high densities ($\sim$100,000 H$_2$ molecules per cc) of molecular cloud cores and protostellar envelopes, a large amount of molecular species (in particular those containing C and O) freeze-out onto dust grain surfaces. It is in these regions that the deuteration of H$_3^+$ becomes very efficient, with a sharp abundance increase of H$_2$D$^+$ and D$_2$H$^+$. The multi-deuterated forms of H$_3^+$ participate in an active chemistry: (i) their collision with neutral species produces deuterated molecules such as the commonly observed N$_2$D$^+$, DCO$^+$ and multi-deuterated NH$_3$; (ii) their dissociative electronic recombination increases the D/H atomic ratio by several orders of magnitude above the D cosmic abundance, thus allowing deuteration of molecules (e.g. CH$_3$OH and H$_2$O) on the surface of dust grains. Deuterated molecules are the main diagnostic tools of dense and cold interstellar clouds, where the first steps toward star and protoplanetary disk formation take place. Recent observations of deuterated molecules are reviewed and discussed in view of astrochemical models inclusive of spin-state chemistry. We present a new comparison between models based on complete scrambling (to calculate branching ratio tables for reactions between chemical species that include protons and/or deuterons) and models based on non-scrambling (proton hop) methods, showing that the latter best agree with observations of NH$_3$ deuterated isotopologues and their different nuclear spin symmetry states.
\end{abstract}


\maketitle

\section{H$_3^+$ and its deuterated forms in dense molecular clouds} \label{sec:htp}
In dark molecular clouds, where interstellar UV photons cannot penetrate, the medium is kept partially ionised by the continuous and unimpeded passage of cosmic rays. This ionisation is crucial for the dynamical \cite{M89} and chemical \cite{HK73} evolution of molecular clouds, which are threaded by magnetic fields and filled with sub-micron-sized dust particles. Cosmic rays ionise the most abundant molecule, H$_2$ (e.g. \cite{SW71,PGG09}), and the main product of this ionisation, H$_2^+$, readily reacts with another H$_2$ molecule to form the most important molecular ion in astrochemistry: H$_3^+$ \cite{W73}, first detected spectroscopically in the laboratory by Oka \cite{O80} and, sixteen years later, in space by Geballe \& Oka \cite{GO96}. H$_3^+$ can easily cede a proton to atoms such as C and O, initiating the path toward molecular complexity \cite{Y17} and it is the main tracer of cosmic-ray ionisation rate in the interstellar medium (e.g. \cite{IM12}).

Typical gas temperatures of molecular clouds are between 10 and 15\,K (e.g. \cite{FP17}) and, in these conditions, the reaction between H$_3^+$ and HD, becomes important \cite{W74,MB89}, producing H$_2$D$^+$ and enhancing the H$_2$D$^+$/H$_3^+$ abundance ratio. More precisely, the following exothermic reactions are at work \cite{PSW92,GHR02}:
\begin{eqnarray}
{\rm pH_3^+ + HD} & \rightarrow & {\rm (o/p)H_2D^+ + (o/p) H_2,} \nonumber \\
{\rm oH_3^+ + HD} & \rightarrow & {\rm oH_2D^+ + (o/p) H_2,} \\
& \rightarrow & {\rm pH_2D^+  + oH_2} \nonumber
\end{eqnarray}
where the symbol "p" or "o" in front of species $i$ indicates the para or ortho form of $i$, while "o/p" means that all combinations (ortho or para) are possible (see Hugo et al. \cite{HAS09} for more details on the exact selection rules). The above reactions are the starting point of deuterium fractionation in cold ($T < 30$\,K) gas, as H$_2$D$^+$ can cede a deuteron to abundant neutral species (e.g. CO and N$_2$), producing deuterated molecular ions (e.g. DCO$^+$ and N$_2$D$^+$) which are crucial tracers of physical conditions in dense and cold regions of interstellar clouds (e.g. \cite{CWZ02,PCP18}), the so-called dense cores \cite{BM89}, within which stars and planets form \cite{BT07,CC12}. If the H$_2$ number densities of dense cores exceed a few $\times$10$^4$\,cm$^{-3}$, CO molecules begins to "catastrophically" freeze-out on dust grains, significantly reducing the CO fractional abundance in the gas phase (e.g. \cite{CWT99}). This implies a reduced destruction rate for both the H$_3^+$ and H$_2$D$^+$ molecular ions, which further boosts deuterium fractionation \cite{DL84}. Indeed, the strongest ground state line of ortho-H$_2$D$^+$ was detected toward one of the the most dynamically evolved and thus centrally concentrated starless core, L1544 in the Taurus Molecular Cloud \cite{CvTC03,CVC08}, where the CO freeze-out and deuterium fractions are also large compared to other less evolved starless cores \cite{CCW05}. The central temperature of L1544 is $\simeq$6\,K \cite{CCW07}. para-D$_2$H$^+$ has also been detected toward two starless cores in the Ophiuchus Molecular Cloud:  16293E \cite{VPY04} and H-MM1 \cite{PBD11}. The importance of these observations is twofold: on the one hand, they point out that multi-deuterated forms of H$_3^+$ have to be included in astrochemical models to reproduce observations (e.g. \cite{RHM03}); on the other hand, they allow us to gain understanding of interstellar chemical processes by providing a test for theoretical predictions. 

The detection of ortho-H$_2$D$^+$ and para-D$_2$H$^+$ in cold regions in the early 2000's prompted the revival of chemical models inclusive of deuterium and spin-state chemistry, first discussed in \cite{PSW92}, as well as fundamental laboratory and theoretical work on state-to-state thermal rate coefficients for reactions of all H$_3^+$ + H$_2$ isotopic variants \cite{HAS09}. The new astrochemical models initially assumed complete freeze-out of species heavier than He \cite{WFP04,FPW04,SHH10}; then complexity increased to include heavier species, in particular N-bearing molecules such as NH$_3$, N$_2$H$^+$ and their deuterated forms \cite{FPW06,PWH09}, which have been found to maintain large abundances toward the CO-depleted regions of starless cores (e.g. \cite{CWZ02,CCW07}). Current chemical models inclusive of spin-state chemistry follow the time dependent evolution of hundreds of species in the gas-phase and on the surface of dust grains (e.g. \cite{SCH13,TCS14}), and even the spin-state chemistry in reactions involving molecules with multiple deuterons \cite{SHC15}, providing detailed predictions for observations.  

\begin{figure}[!h]
\centering\includegraphics[width=\textwidth]{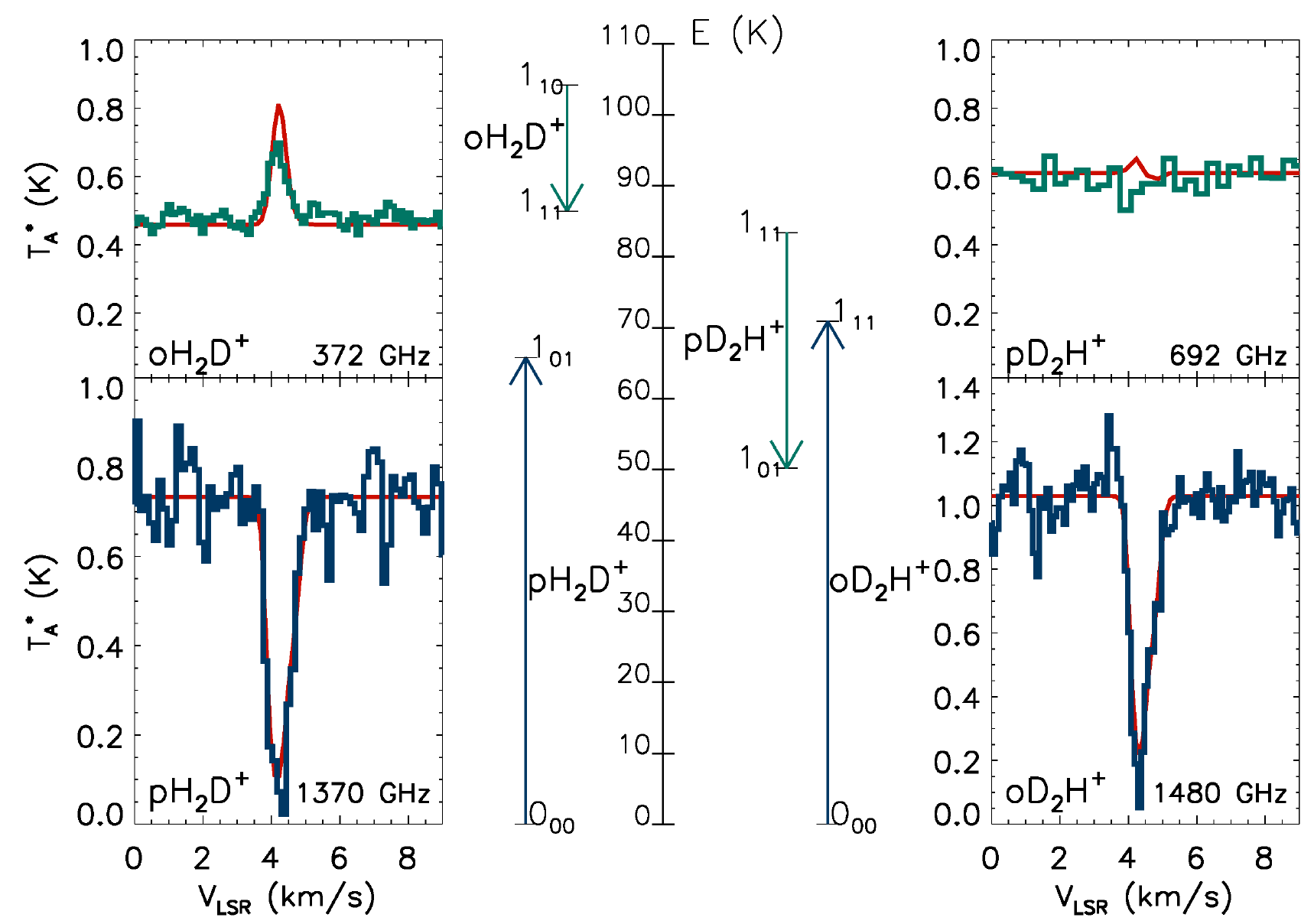}
\caption{Spectra of ortho-H$_2$D$^+$, para-H$_2$D$^+$, para-D$_2$H$^+$ and ortho-D$_2$H$^+$ toward IRAS 16293-2422 (histograms) as observed by  Br\"unken et al. \cite{B14} and Harju et al. \cite{H17}. The red curves are the lines predicted from the radiative transfer applied to a model cloud with number density and temperature profiles appropriate for IRAS 16293-2422 and with molecular abundance profiles predicted by the Sipil\"a et al. \cite{SCH13,SHC15} chemical codes, where spin-state chemistry is followed in detail. The energy levels in the centre of the figure schematically show the four ground-state transitions of the four H$_3^+$ isotopologues.}
\label{fig:htp_family}
\end{figure}

The first detailed comparison between model predictions for the various H$_3^+$ isotopologues and observations has been carried out in the direction of a very young stellar object, still embedded in the parent cloud with physical and chemical characteristics similar to those of evolved starless cores (IRAS 16293-2422; \cite{B14,H17}). The ortho-H$_2$D$^+$ and para-D$_2$H$^+$ ground state lines have frequencies centred at 372.4\,GHz and 691.7\,GHz, respectively, and they can be both observed with the Atacama Pathfinder Experiment (APEX). The ground-state lines of para-H$_2$D$^+$ at 1.37\,THz and of ortho-D$_2$H$^+$ at 1.48\,THz, required the Stratospheric Observatory for Infrared Astronomy (SOFIA) to be detected.  Br\"unken et al. \cite{B14} used APEX and SOFIA to detect both the ortho and para forms of H$_2$D$^+$ toward IRAS 16293-2422. This provided the first detection of para-H$_2$D$^+$ at 1.37\,THz and the first measurement of the ortho-to-para ratio of H$_2$D$^+$, linked to the ortho-to-para ratio of H$_2$ \cite{HAS09}, important to put constraints on the cloud chemical age. Toward the same source, Harju et al. \cite{H17} also detected for the first time ortho-D$_2$H$^+$ at 1.48\,THz with SOFIA. Using the chemical models from Sipil\"a et al.  \cite{SCH13,SHC15}, coupled with the physical structure of the source, the predicted and observed line intensities of all the H$_3^+$ isotopologues observed toward IRAS 16293-2422 are reported in Figure\,\ref{fig:htp_family}, showing the good agreement between models and observations. This study also established that rotational excitation of the reactant species needs to be taken into account when computing rate coefficients, particularly important for protostellar envelopes such as that surrounding IRAS 16293-2422 \cite{H17,SHC17}, typically denser and slightly warmer than starless cores. 

The large abundance of deuterated H$_3^+$ isotopologues in cold and dense gas also accounts for the predicted large atomic D/H ratio in the gas phase, as a copious amount of D atoms is produced by the dissociative recombination of H$_2$D$^+$, D$_2$H$^+$ and D$_3^+$. Large D/H ratios in cold and dense regions, where CO molecules are mainly frozen onto the surface of dust grains, lead to the huge (13 orders of magnitude!) deuterium fraction measured in CH$_3$OH, as D atoms compete with H atoms in the deuteration/hydrogenation of solid CO (e.g. \cite{CC12,PCH04}), and explains the different levels of D-fractions measured in organics and water, similar to that found in our Solar System \cite{CCB14,E02}. A detailed knowledge of the chemistry regulating deuterium fractionation is important to be able to use molecules as powerful diagnostic tools in star and planet forming regions, and to unveil the chemical/dynamical history of our Solar System. 



\section{Deuteration of ammonia in quiescent dense clouds} \label{sec:sipila} 

As already mentioned in Section\,\ref{sec:htp}, ammonia is a great tracer of dense and cold clouds (e.g. \cite{BM89}). Ammonia deuteration follows that of H$_3^+$. Singly, doubly and triply deuterated forms of ammonia have been detected in space (e.g. \cite{RLvT05} and references therein) and they are known to be good tracers of physical conditions in dense clouds, preferentially tracing the cold and CO-depleted centres of evolved starless cores (e.g. \cite{CCW07}), where future stellar systems will form. Sipil\"a et al. \cite{SHC15} built a chemical model where all the deuterated forms of ammonia and their spin states are followed as a function of time and physical conditions. In this model, the different nuclear spin symmetry states of molecules containing two or more H and/or D nuclei are treated as separate species. Symmetry rules are applied in the context of the complete scrambling assumption to calculate branching ratio tables for reactions between chemical species that include multiple protons and/or deuterons (up to 6 H/D atoms), following permutation symmetry algebra \cite{Q77,HAS09}. The model predictions showed that the abundances of the various NH$_3$ deuterated isotopologues, the ortho-to-para abundance ratios of NH$_2$D and NHD$_2$, the meta-to-ortho and meta-to-para abundance ratios of ND$_3$ are very sensitive to variations in volume densities and temperatures; thus, one could use observations of the above species to put stringent constraints on the cloud physical structure. At the same time, if the physical conditions of a certain cloud or dense core have been determined with other methods, the observations of the above-mentioned species could be used to compare with model predictions and test the basic assumptions of our chemistry. 

In order to test our current understanding of spin-state chemistry of deuterated species, Harju et al. \cite{H17b} measured the abundances and spin ratios of NH$_2$, NHD$_2$ and ND$_3$ toward a well-known starless core in the Ophiuchus Molecular Cloud, HMM1. These observations found large D-fractions, with NH$_2$D/NH$_3$ $\simeq$ 0.4, NHD$_2$/NH$_2$D $\simeq$ 0.2 and ND$_3$/NHD$_2$ $\simeq$ 0.06, which can be roughly (within a factor of less than 2) reproduced by the chemical model after 3$\times$10$^5$\,yr. However, the model predicts too low ortho-to-para-NH$_2$D ratios and too large ortho-to-para-NHD$_2$ ratios, which are measured to be close to their corresponding nuclear spin statistical weights. This disagreement between model predictions and observations indicates that our basic assumption of complete scrambling may not be correct. Indeed, Le Gal et al. \cite{LGX17} recently performed an observational and theoretical study of H$_2$Cl$^+$, finding that its major formation reaction (HCl$^+$ + H$_2$ $\rightarrow$ H$_2$Cl$^+$ + H) produces this ion via a hydrogen abstraction rather than a scrambling mechanism. Inspired by these recent findings, we modified our model to eliminate the full scrambling assumption from the Sipil\"a et al. \cite{SHC15} model and include instead the proton hop mechanism. The new model is applied to proton-donation reactions only. The results of this new work are presented in the next section and will be fully described in Sipil\"a et al. (in prep.). 

\section{Scrambling or proton hop?}
Before trying to answer the question of this section, we review the basic assumptions behind the full scrambling and proton hop mechanisms. 

Full-scrambling branching ratios are calculated as follows. We consider that reactions proceed through an intermediate complex, which constitutes a pool of atoms from which the atoms are drawn one by one, and the rate coefficient is constructed by multiplying together simple probabilities. For example, in the reaction $\rm NH_3 + D_2H^+ \longrightarrow (NH_4D_2^+)^* \longrightarrow NH_3D^+ + HD$, the intermediate complex contains four H atoms and two D atoms. The four H/D atoms to create $\rm NH_3D^+$ can be picked in ${4 \choose 1} = 4$ equivalent combinations: HHHD, HHDH, HDHH, and DHHH. The individual probabilities for each combination are the same, for example for HHHD: $\frac{4}{6} \frac{3}{5} \frac{2}{4} \frac{2}{3} = \frac{2}{15}$. So, in total, the branching ratio for $\rm NH_3D^+$ formation is ${4 \choose 1} \frac{2}{15} = \frac{8}{15}$. There is only one way to form $\rm NH_4^+$ (probability $\frac{1}{15}$), and so the probability to form $\rm NH_2D_2^+$ is $\frac{6}{15}$.

\begin{figure}[!h]
\centering\includegraphics[width=\textwidth]{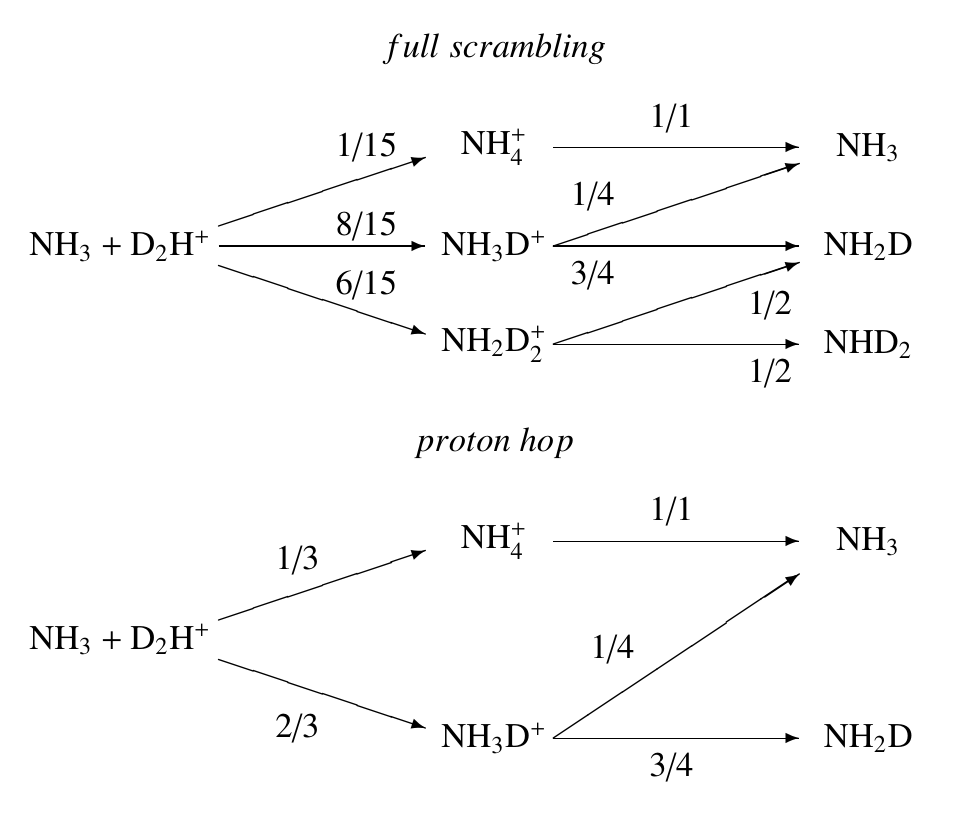}
\caption{Simplified branching ratio diagrams, showing NH$_3$ formation only, for the case of the proton hop mechanism (left diagram) and full scrambling (right diagram).}
\label{fig:scrambling}
\end{figure}

The proton hop mechanism instead assumes that the donating ion simply loses one of its atoms, and the reaction as a whole does not go through an intermediate complex -- at least as far as the branching ratio calculation is concerned. Therefore the probability to form $\rm NH_3D^+$ in $\rm NH_3 + D_2H^+$ is simply $\frac{2}{3}$, for example. We obtain in general fewer formation pathways when assuming a proton/deuteron hop instead of full scrambling. Figure\,\ref{fig:scrambling} illustrates the situation.
 
\begin{figure}[!h]
\centering\includegraphics[width=\textwidth]{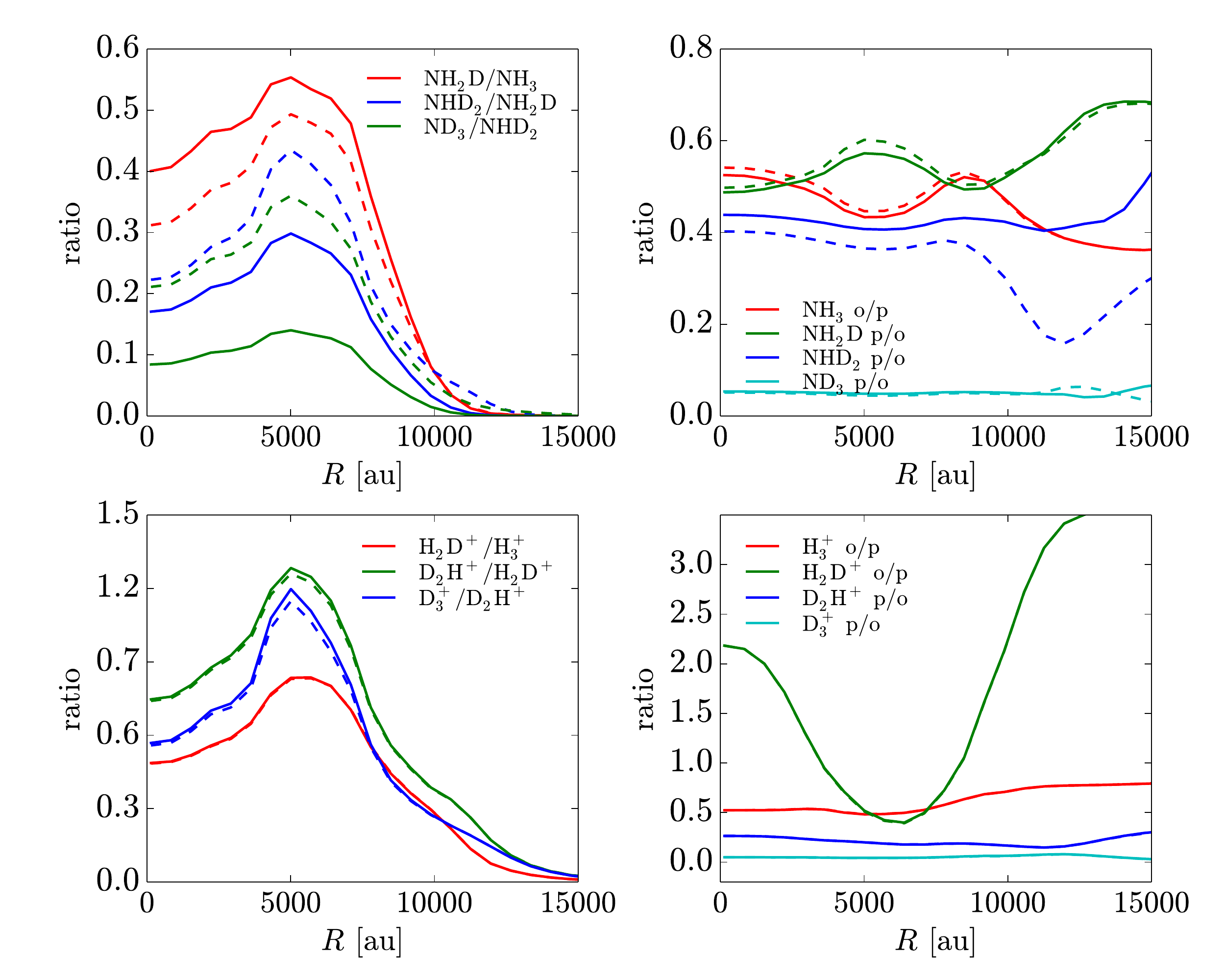}
\caption{Comparison between model predictions from chemical models assuming full scrambling (dashed curves) and proton hop (solid curves) based on Sipil\"a et al. \cite{SHC15} and Sipil\"a et al. (in prep.), respectively. The abundance ratios are plotted as a function of the distance (in astronomical units) from the core centre. The left panels show deuterium fractions of ammonia (top) and H$_3^+$ (bottom) isotopologues, while the right panels show the nuclear spin-state ratios of ammonia (top) and H$_3^+$ (bottom) isotopologues. Significant changes are only present in the case of NH$_3$ species.}
\label{fig:comparison}
\end{figure}

We note that these approaches to calculating the branching ratios are extremely simplistic, and are based purely on probabilistic considerations. They neglect, for example, any zero-point energy effects that come into play when introducing different counts of hydrogen or deuterium in the reactions. Nevertheless, we decided to test if proton hop, instead of full scrambling, could improve the agreement between model predictions and the observations of Harju et al. (2017b), described in Section\,\ref{sec:sipila}.  With this aim, we run two different chemical models of the same molecular cloud core, with physical structure given by Keto et al. \cite{KRC14}\footnote{Although the physical structure derived by Keto et al. \cite{KRC14} refer to the dynamically evolved starless core L1544 in Taurus, we note that L1544 and HMM1, the core studied in Harju et al. \cite{H17b}, have similar physical structure as well as similar  levels of deuteration.}: one assuming full scrambling, based on Sipil\"a et al. \cite{SHC15}, and one assuming proton hop, based on Sipil\"a et al. (in prep.). Figure\,\ref{fig:comparison} shows the results of this comparison, with each panel reporting abundance ratios as a function of cloud radius. Because of the centrally concentrated structure of the core (with central H$_2$ number densities of about 10$^{6}$\,cm$^{-3}$ within a radius of $\simeq$2000\,au \cite{CPC19}) the various molecules in Fig.\,\ref{fig:comparison} mainly trace the inner few thousand au, so the central values should be considered for the comparison with observations toward the core centre.  The first thing to note is that significant differences are present between scrambling and proton-hop models in the case of NH$_3$ isotopologues, while the H$_3^+$ isotopologues are not affected by the mechanism change. In particular: (1) the deuteration ratios within the central few thousand au predicted by the proton-hop model are in very good agreement with the observed values ($\simeq$0.4, 0.2 and 0.06 for NH$_2$D/NH$_3$, NHD$_2$/NH$_2$D, and ND$_3$/NHD$_2$, respectively; see Sect.\,\ref{sec:sipila}); (2) the ortho-to-para ratios of NH$_2$D and NHD$_2$ are also in better agreement with observations, with the first one increasing and the second one decreasing, as required by the analysis of Harju et al. \cite{H17b}. Although a proper line radiative transfer analysis needs to be carried out to quantify the agreement between model predictions and observations, overall we can conclude that proton hop appears to be a better representation of nuclear spin-state chemistry, in agreement with the conclusion of Le Gal et al. \cite{LGX17}. 

As a final note, it is important to point out that Hily-Blant et al. \cite{HFR18}, considering the observational uncertainties, found satisfactory agreement between the observed spin and deuteration ratios and those predicted by their combined dynamical and chemical model where full scrambling was assumed. However, the full scrambling mechanism, which depends on the lifetime and the structure of the reaction complex can also be questioned on theoretical grounds. Currently we are missing theoretical and experimental data which could be used to decide if it is the proton hop or the full scrambling scenario which better describes the formation of interstellar ammonia and its deuterated isotopologues. 

\section{Conclusion}
H$_3^+$ and its deuterated forms are crucial ingredients in the chemical structure and evolution of dense cloud cores on the verge of star formation and of protostellar envelopes. They affect the abundance and deuteration ratio of important tracers of dense and cold material, such as NH$_3$ and N$_2$H$^+$.  Therefore, comprehensive chemical models, inclusive of deuterium and nuclear spin-state chemistry, are needed to interpret observations and also to test basic assumptions behind the theory. Here we have reviewed theoretical and observational work on spin chemistry toward dense clouds cores and young protostars, showing that a combination of ground-based (such as APEX and IRAM) and space-borne (SOFIA) telescopes can provide detailed information about cloud ages and the (unobservable-in-cold-clouds) ortho-to-para H$_2$ ratio, which strongly affects deuterium fractionation. From the comparison between model predictions and NH$_3$ isotopologue observations toward a cold and quiescent dense core, we arrived at the conclusion that the basic assumption of full scrambling, adopted in our comprehensive chemical models inclusive of deuteron and nuclear-spin chemistry, may not be correct. To test this point, we built new chemical models of dense cloud cores, presented here for the first time, where full scrambling has been substituted by the proton hop mechanism, and compared the full scrambling with the proton-hop models as well as with observational results. The main result is that indeed the proton-hop mechanism better reproduces the observations of NH$_3$ isotopologues and their different nuclear spin symmetry states, although more experimental and theoretical work is needed to arrive at a definitive conclusion. In the future, with the use of our radiative transfer code, model line profiles will be simulated and compared with the observed spectra, to test further our models and start finally to use our predictions to constrain the physical and chemical structure of star forming dense clouds, the first steps toward stellar systems like our own. 

\vskip6pt

\enlargethispage{20pt}



\aucontribute{PC wrote the paper and participated in the interpretation of observations and comparison with chemical models. OS developed the comprehensive chemical models presented in this article and helped with the paper writing. JH carried out observations and actively helped in the developments of the  chemical models and the paper writing.}


\funding{Max Planck Society}

\ack{The authors thank Sandra Br\"unken and Stephan Schlemmer for their continuous help with observations and nuclear spin theory.}



\end{document}